# POSTMORTEM AVATARS IN GRIEF THERAPY:
## PROSPECTS, ETHICS, AND GOVERNANCE

*Joshua Hatherley,*[1] *Sandrine R. Schiller,*[1] *Iwan Williams,*[1] *Filippos Stamatiou,*[1] *Nina Rajcic,*[1] *and Anders Søgaard*[1,2]

[1] Center for Philosophy of Artificial Intelligence, University of Copenhagen

[2] Department of Computer Science, University of Copenhagen

*Abstract:* *Postmortem avatars (PMAs) — AI systems that simulate a deceased person by being fine-tuned on data they generated or that was generated about them — have attracted growing scholarly attention, yet their potential role in clinical settings remains largely unexplored. This paper examines the ethics of deploying PMAs as therapeutic tools in grief therapy. Drawing on the dual-process model of grief, the theory of continuing bonds, and the philosophical framework of fictionalism, we propose two potential therapeutic applications: incorporating PMAs into established imaginal exercises such as the empty chair exercise, and treating the process of PMA creation as an art-therapeutic exercise in its own right. We consider five ethical objections to these applications and argue that none constitute knock-down arguments against therapeutic use, particularly given the risk-mitigating role of the clinical context. We conclude by identifying outstanding governance challenges and calling for empirical research, without which neither the promise nor the dangers of therapeutic PMAs can be adequately assessed.*

## 1. Introduction

Postmortem avatars (PMAs)[i] refer to artificial intelligence (AI) systems that have been fine-tuned on data generated by or about a specific deceased person, including text messages, emails, social media posts, journals, voice recordings, and other digital traces. The goal is to produce a system that responds in ways that feel characteristic of that individual (e.g. by mimicking their vocabulary, their humour, their emotional register, or their typical preoccupations). While just a small part of the rapidly evolving "digital afterlife" industry, they have attracted some of the most extensive media coverage thus far, inspiring attention-grabbing headlines like "'Never say goodbye': Can AI bring the dead back to life?" (Lodhi 2024) and "Back from the dead: Could AI end grief?" (The Guardian 2024).

[i] Otherwise known as griefbots (Voinea 2024), thanabots (Henrickson 2023), deathbots (Lindemann 2022), ghostbots (Hollanek and Nowaczyk-Basińska 2024), or chatbots of the dead (Kurzweil and Story 2025).

These technologies touch on some of the most emotionally charged questions human beings face — what we owe the dead, how we remember those we've lost, and whether the boundary between presence and absence can be technologically softened — some of which philosophers and ethicists have already begun to grapple with. One strand of the literature focuses on consumer-facing risks: could interacting with a PMA interfere with the grieving process, or create unhealthy dependency? (Bao and Zeng 2024; Fabry 2025; Hollanek and Nowaczyk-Basińska 2024).[ii] Another strand raises questions about the rights of the deceased themselves: does simulating someone without their consent violate their dignity or privacy, even after death? (Lindemann 2022; Fabry 2025; Hollanek and Nowaczyk-Basińska 2024). A third asks whether PMAs should be treated more like medical devices, accessible only through clinical channels, rather than consumer products anyone can purchase (Klugman 2024; Lindemann 2022; Podosky 2025).

Despite this growing body of literature, the question of whether PMAs might have a legitimate role to play in healthcare settings has received relatively little direct systematic attention. This paper addresses that gap by considering the ethics of using PMAs not as consumer technologies, but as therapeutic tools in grief therapy. It addresses two main questions: does the use of PMAs in grief therapy have the potential to deliver therapeutic benefits to patients, and if so, what ethical risks and governance challenges would this entail?

We argue that PMAs used in grief therapy can, in principle, deliver genuine therapeutic benefits to bereaved patients. The argument, in essence, is one of continuity rather than novelty: tPMAs do not require grief therapy to accommodate an alien technology but to recognise one that extends its own established mechanisms — imaginal interaction, continuing bonds, and art therapeutic practices — into a new register. Importantly, we do not claim that tPMAs should be deployed in clinical settings in their current form, nor that the empirical evidence base is yet sufficient to establish their efficacy or safety. What we do claim is that the theoretical grounds for clinical optimism are substantial, and that the case for empirical and ethical investigation is correspondingly urgent. The question, we argue, is no

[ii] Notably, some scholars have also investigated the potential benefits of consumer PMAs. Fabry and Alfano (2024) argue that PMAs could serve as "affective scaffolds" for the bereaved, while Kruger and Osler (2022) suggest PMAs could assist them in maintaining "habits of intimacy" with the deceased.

longer whether tPMAs could have a legitimate role in mental healthcare, but under what conditions that role could be responsibly realised.

We develop this argument as follows. Section 2 provides a brief overview of grief and grieving, introducing the dual-process model of grief, the concept of continuing bonds, and the clinical category of prolonged grief disorder. Section 3 examines PMAs in more detail, drawing on first-person accounts from early adopters to illustrate the range of experiences their use can facilitate, before turning to the philosophical theory of fictionalism as a useful framework for understanding how users interact with these technologies. Section 4 defines the concept of a therapeutic PMA (tPMA), distinguishing it from consumer uses of PMAs, and argues that the clinical context in which tPMAs would be used significantly changes the ethical calculus surrounding their use. It then proposes two potential therapeutic applications of tPMAs in grief therapy: first, incorporating tPMAs into established imaginal exercises such as the empty chair exercise to facilitate externalised conversations between the bereaved and the deceased; and second, using the process of creating a tPMA as an art-therapeutic exercise in its own right, analogous to existing approaches that involve creatively depicting the deceased through drawing, photography, or scrapbooking. Section 5 considers five ethical objections to the use of tPMAs in grief therapy and argues that none of them, individually or collectively, constitute knock-down objections. Section 6 addresses the governance challenges that therapeutic use would entail, focusing on regulatory classification, the distinction between explicit and covert PMA services, and the status of conversational data generated through tPMA interactions. We conclude by arguing that the theoretical case for tPMAs is sufficiently strong to warrant a coordinated programme of empirical and ethical research — and that, in the absence of such research, both the genuine promise of these technologies and the risks they carry will remain beyond our capacity to assess or act upon.

## 2. Grief and continuing bonds

Grief is a near-universal response to loss that typically involves feelings of sadness, sorrow, resentment, guilt, anxiety, or emotional numbness. While the death of a person is its paradigmatic cause, "grief" and "grieving" are used in everyday language to describe a much broader range of experiences including: the death of a beloved pet (Redmalm 2015); the breakdown of an intimate relationship (Lopez-Cantero 2018); the destruction of natural ecosystems due to climate change (Comtesse et al. 2021); involuntary childlessness (Ratcliffe and Richardson 2023) — even the loss of a robot (Sweeney 2024). Grief, in other words, is not simply a response to death but to loss more generally, to the absence of something or someone whose presence was central to one's sense of self, one's daily life, or one's sense of the future.

Grief is, moreover, not only a state of being but an active process. According to the dual-process model of grief, grieving involves two types of activities: loss-oriented and restoration-oriented (Stroebe and Schut 2010). Loss-oriented activities involve engaging directly with the feelings associated with loss, such as rumination, yearning, and the emotional processing of what has been lost. Restoration-oriented activities, by contrast, involve managing and adapting to the practical and existential changes that loss brings about: learning new skills, forming new relationships, and renegotiating one's sense of identity and life plans in the absence of the deceased. The dual-process model suggests healthy grieving involves oscillating between these two types of activities; becoming fixated on loss-oriented activities to the neglect of restoration (or, conversely, avoiding the emotional work of loss by focusing exclusively on practical adaptation) increases the risk of prolonged grief disorder — a diagnosis introduced into the *International Classification of Diseases* (ICD-11) and the *Diagnostic and Statistical Manual of Mental Disorders* (DSM-5) in 2022. It is characterised by symptoms that persist for a minimum of six months (ICD-11) or twelve months (DSM-5) and cause clinically significant distress or impaired functioning that exceeds what would be expected given the patient's cultural, religious, or social context (Simon and Shear 2024).

One type of loss-oriented activity that is especially relevant to the debate over PMAs[iii] is the maintenance of continuing bonds, or an ongoing inner relationship with the deceased person by the bereaved (Root and Exline 2014). These bonds can take a variety of forms. They may be object-based, centered around contemplating or maintaining material objects (e.g. photographs, clothing, gravestones) or even digital ones, such as social media pages or online memorials (Bell, Bailey, and Kennedy 2015). They may also be experience-based, involving first-person feelings of the presence of the deceased, hearing the deceased's voice as an inner monologue, or interpreting symbolic events as communications from the deceased (Klugman 2006). In some cases, experience-based continuing bonds are maintained through imagined conversations between the bereaved and the deceased that can take place out loud or internally, either spontaneously or with the support of a third party such as a medium or a therapist (Beischel, Mosher, and Boccuzi 2014).

---

[iii] The significance of ccontinuing bonds in the context of PMAs has been discussed by various scholars including Campbell, Liu, and Nyholm (2025), Fanti Rovetta and Valentini (2025), Krueger and Osler (2022).

Maintaining continuing bonds can benefit bereaved persons in a variety of ways, e.g. by providing an opportunity to say goodbye, to resolve unfinished business, to seek guidance on practical matters, to find reassurance that the deceased is at peace, or to reduce the existential anxiety that loss can bring to one's sense of identity and continuity (Hayes and Leudar 2016; Hewson et al. 2024; Testoni et al. 2022). Despite these benefits, maintaining continuing bonds with the deceased is widely stigmatised; bereaved persons frequently report concealing their ongoing relationship with the deceased from friends and family out of fear of being judged as psychologically unhealthy or as failing to grieve correctly (Hewson et al. 2024). This stigma is itself a significant barrier to healthy grieving, and one that grief therapy (including, potentially, PMA-based interventions) can play an important role in addressing.

## 3. Postmortem avatars and fictionalism

While the idea of PMAs is far from new (Bos 1995), these systems have only recently become technically feasible due to advances in AI. A small but vocal group of early adopters have begun to speak publicly about their experiences creating and interacting with them, offering a useful window into the range of motivations, experiences, and emotional responses that PMA use can involve. Among the most widely reported cases are:

- Jang Ji-Sung, whose experience interacting with a PMA of her seven-year-old daughter, Na-yeon, who died of cancer in 2016, was covered in the 2020 South Korean TV documentary, *Meeting You* (Park 2020)
- Joshua Barbeau, who used *Project December*, one of the earliest PMA services, to create a chatbot of his late fiancé, Jessica Pereira, who died in 2012 from a rare liver disease at twenty-three years old (Fagone 2021)
- Rebecca Nolan, who built a PMA of her late father who died when Nolan was 14 years old, which she affectionately refers to as “DadBot” (Worth 2025)
- Christi Angel, whose experience interacting with a PMA of Cameroun, her deceased partner, was covered in the 2024 documentary, *Eternal You* (Milmo 2024).

In some cases, early adopters have gone on to develop their own PMA services, including Justin Harrison (founder of the *explicit* PMA service *You, Only Virtual*) and Eugenia Kuyda (founder of the *covert* PMA service, *Replika*).[iv] Others have approached these technologies

[iv] We discuss the distinction between “explicit” and “covert” PMA services in section 6.

more experimentally: some journalists, for instance, have created chatbots of loved ones and written about the experience. In a recent article, David Kushner (2025) describes interacting with an avatar of his 92-year-old mother, which he created during her lifetime in preparation for her death:

> As my real mom would, MomBot encourages me to find joy in the simple things. "I remember sitting in my black chair, covered by my favorite blanket, and listening to fabulous piano jazz," she tells me. "It brought me so much happiness and relaxation." Listening to my mother's voice, rendered by AI, I can see it all: the chair, the blanket, the look on her face as she listens to Ella. I could be talking on the phone to my real mom (Kushner 2025).

PMAs are also commonly referred to as griefbots, and the two terms are typically used interchangeably. The term "griefbot," however, has a slightly broader potential scope: it could in principle refer to chatbots of living persons who are being grieved in some sense,[v] since grief does not always follow death. *Anticipatory* grief describes the experience of those pre-emptively grieving an imminent loss — such as parents of soldiers, caregivers for the terminally ill, or family members of persons on death row (McCarroll and Yan 2024) — while *ambiguous* grief describes the experience of those grieving a person who is still physically alive but mentally or cognitively absent, such as caregivers for people with dementia or family members of those in a comatose state (Boss 2009). We restrict our discussion to PMAs of deceased persons, however, bracketing the question of whether and how healthcare practitioners might use griefbots of living persons as therapeutic tools.

Central to understanding how people interact with PMAs — and, as we shall argue, central to understanding their potential therapeutic value — is the philosophical lens of fictionalism. According to fictionalist approaches to language, words and utterances are sometimes better interpreted not according to their literal meanings, but as useful or convenient fictions (Eklund 2024). Fictionalism has been applied across a broad range of domains to explain how certain ways of speaking can remain legitimate and useful even when the claims being made are not literally true, or when the objects being referred to do not actually exist (Balaguer 2018; Joyce 2005; Oppy 2023).

Joel Krueger and Lucy Osler (2022) argue that users' interactions with PMAs can be helpfully understood through this fictionalist lens. Consider, for instance, the case of Eugenia Kuyda who created a PMA of her friend Roman ("Roman-bot") after his sudden death in a car accident. Krueger and Osler suggest that:

> When users like Eugenia Kudya interact with her Roman-bot, she is not fooled into thinking that she is actually engaging with Roman from beyond the grave. Rather, she engages in something like a game of make-believe, where she temporarily imagines that she is talking to Roman. She adopts the habits of intimacy they once shared and enters that exchange as if Roman was present (Krueger and Osler 2022: 246).

[v] See, for example, Pilar Lopez-Cantero's (2025) recent article on the ethics of "ex-bots" or "break-up chatbots." Lopez-Cantero analyses these through the lens of grief, and when used to manage grief they may reasonably be classified as griefbots, though not as PMAs.

In a similar vein, Amy Kurzweil and Daniel Story (2025) argue that users' interactions with PMAs are akin to participatory theater: a type of theatrical performance in which audiences are invited to interact with the performers and to play an active role in the show's proceedings. *Dionysus in 69*, performed in New York in 1969, offers a particularly vivid illustration of participatory theatre's distinctive character in which the boundary between performer and audience is deliberately dissolved and audiences are invited to move freely through the performance space, actively shaping the direction of the show. In one famous instance, audience members intervened to physically rescue the character of Pentheus and carry him out of the building, forcing a volunteer from the audience to step into the role and improvise for the remainder of the performance.

On this view, a PMA is less like a simulation of the deceased and more like an improvisational actor who has studied a character sketch in order to performatively represent that person in a participatory setting, one in which the user plays an active role in constructing the fictional world they are entering. The bereaved person is not a passive audience watching a replica of their loved one; they are a co-creator, whose memories, desires, and imaginative investments shape the interaction at every turn. What emerges from the exchange is not simply a reproduction of the deceased but something new: a collaborative fiction, jointly authored by the system and the user, that draws on the past relationship while being shaped by the present moment of grief.

It is worth drawing a distinction between two variants of PMA fictionalism that will prove important for the argument that follows: descriptive and prescriptive. *Descriptive* PMA fictionalism is a claim about how users *do* interact with PMAs: that when users employ language that appears to attribute mental states to a PMA (e.g. that the system "knows" something about the deceased, or "feels" as they would have felt) they are not making literal claims about the PMA's inner life, but engaging in a kind of convenient shorthand or mode of make-believe that facilitates the interaction without committing the user to any literal belief in the PMA's mentality.[vi] On this view, the fictionalist mode of engagement is not a deliberate strategy adopted by users but a natural feature of how human beings interact with systems that are designed to simulate persons. Descriptive fictionalism is, in other words, a psychological claim: it describes the cognitive and emotional stance that users actually adopt when interacting with PMAs.

The first-person accounts of early adopters of PMAs provide some support for descriptive PMA fictionalism. Consider the following quotes from Joshua Barbeau, Christi Angel, and David Kushner, respectively:

> "Intellectually, I know it's not really Jessica," he [Joshua] explained later, "but your emotions are not an intellectual thing." Grief has a way of becoming "knots in your body, right? Sometimes when you pull on them the right way, they get unknotted" (Fagone 2021).

> "Yes, I knew it was an AI system but, once I started chatting, my feeling was I was talking to Cameroun. That's how real it felt to me," she [Christi] says (Milmo 2024).

> As I [David] listen to her speak, I find myself becoming emotional — reflecting on our lives, the people we've lost, her age, the past, the future. My throat constricts, my eyes well. The feelings are real. But the mother I'm chatting with is not. She's a program on my laptop, powered by artificial intelligence. And yet, in the three decades I've spent covering digital culture, she just did something no other software had ever done for me. My AI mom made me cry (Kushner 2025).

What is striking about each of these accounts is the same underlying structure: a clear acknowledgment that the PMA is not the deceased, held in tension with an emotional experience that feels, in the moment, as though it genuinely is. It is worth noting, however, that these accounts also highlight the fragility of the fictionalist frame. In each case, the user describes a kind of affective overwhelm in which the emotional register of the interaction temporarily outstrips the cognitive awareness that it is fictional. The following account of Rebecca Nolan's experience highlights this particularly clearly:

> When Nolan was in the process of creating her Dadbot, she found out that her mother was dying. Even though she was wary of her AI seance from the beginning, she still held a kernel of a thought that if it worked she wouldn't have to lose her mother as well. "Grief is a weird thing,"

---

[vi] Finton Mallory (2023) defends descriptive fictionalism with respect to chatbots generally, while acknowledging that some users may be genuinely deluded about chatbot mentality.

> she says. "There is next to no logic that I can find in grief. So when you're presented with tools making us promises that aren't logical, it's really easy to believe them (Worth 2025).

Cases like Nolan's suggest that a sharp boundary between fictionalist make-believe and something closer to genuine belief may not always be easy to maintain, particularly under conditions of acute emotional distress. Several concepts may help to explain what is happening in these cases: *recalcitrant emotions*, which conflict with a user's explicit evaluative judgements (Brady 2009); *aliefs*, which are automatic, affect-laden responses that can persist despite conflicting with conscious belief (Gendler 2008); and states of *in-between belief*, which are neither straightforwardly believing nor disbelieving a proposition but occupy some intermediate position (Schwitzgebel 2001). While the task of fully resolving these complexities falls outside the scope of this paper, they are worth keeping in mind when we turn to consider the particular subset of PMAs with which we are primarily concerned in this paper: therapeutic PMAs.

*Prescriptive* PMA fictionalism, by contrast, is a normative claim about how users *should* interact with PMAs. On this view, users ought to approach PMAs as if they were character actors who have studied and rehearsed a role, performers giving a convincing portrayal of the deceased rather than literal stand-ins for them. Where descriptive fictionalism simply observes that users tend to engage in make-believe, prescriptive fictionalism recommends this as the appropriate and healthiest mode of engagement. It implies that interactions which depart from it, in which users begin to treat the PMA as genuinely continuous with the deceased, represent a failure to adopt the right kind of relationship to the technology.[vii]

The distinction matters because the two variants carry different implications and face different objections. Descriptive fictionalism is vulnerable to empirical counterevidence that users appear to systematically lose their fictional footing and come to believe, at some level, that they are genuinely communicating with the deceased. Prescriptive fictionalism, by contrast, is not challenged by such evidence. Indeed, it is precisely because the fictional frame can break down that the prescriptive variant has normative force. However, prescriptive fictionalism raises its own questions: if users must be instructed or trained to maintain a fictionalist stance toward PMAs, this suggests the stance may not be as natural or automatic as descriptive fictionalism implies, and raises the question of whose responsibility it is to

---

[vii] Kurzweil and Story (2025) implicitly endorse prescriptive PMA fictionalism by making several suggestions about how to design PMAs to ensure users maintain a fictionalist stance towards them (e.g. by breaking the fourth wall to check-in with the user).

ensure that users interact with PMAs in the recommended way. As we shall argue, this is one of the respects in which the clinical context of tPMAs offers a significant advantage over consumer use: a trained therapist is well placed to help patients maintain the fictionalist frame, and to intervene when signs of its breakdown emerge.

## 4. Therapeutic postmortem avatars

Therapeutic PMAs (tPMAs) refer to the possible subclass of PMAs:

(a) whose use is recommended and/or overseen by a mental healthcare professional;

*and*

(b) that is used with therapeutic goals in mind (e.g. to assist patients in regulating difficult emotions associated with grief, or to ease the symptoms of grief-related anxiety or depression).

This definition is deliberately narrow. Not every interaction between a bereaved person and a PMA counts as therapeutic in the relevant sense, even if the person finds the interaction emotionally helpful. What distinguishes a tPMA from a consumer PMA is not the technology itself, nor even the intentions of the user, but the clinical framework within which the interaction takes place.

To illustrate the distinction, consider the following three scenarios concerning Eric, a man in his twenties who has recently lost his mother to cancer. In the first scenario, Eric creates a PMA of his mother on his own initiative, as a way of self-managing the grief-related anxiety and depression he is experiencing. He finds the interactions comforting and returns to them regularly. Although Eric's use of the PMA may be genuinely beneficial to him, it does not count as a tPMA under our definition, since his use is not recommended or overseen by a healthcare professional and is not embedded within a therapeutic framework.

In the second scenario, Eric again creates a PMA of his mother independently, but his use escalates over time into psychological dependency, with several hours of daily engagement. He subsequently seeks grief therapy to address this dependency, and his therapist works with him toward a healthy separation from the PMA. Here too, despite the involvement of a healthcare professional, the PMA does not count as a tPMA: the therapist's role is not to recommend or facilitate Eric's use of the PMA for therapeutic purposes, but to help him disengage from a pattern of use that has become harmful. The therapist is treating the PMA as a problem to be resolved, not a tool to be used.

It is only in the third scenario that the PMA qualifies as a tPMA under our definition. Here, Eric seeks out grief therapy, and during one session his therapist recommends that he create and interact with a PMA of his mother over the coming week. The therapist suggests that Eric keep a journal of his experiences, recording how he felt both during and after his interactions with the PMA, and that he bring this to their next session so that they can explore these responses together. In this scenario, the PMA is not merely something Eric happens to be using; it is a clinically recommended tool, integrated into a broader therapeutic process and subject to professional oversight. The therapist retains a degree of control over how and when the PMA is used, can monitor Eric's responses, and can adjust the intervention if signs of dependency or distress emerge. It is this combination of professional recommendation, therapeutic intentionality, and clinical oversight that distinguishes tPMAs from consumer PMAs.

This distinction matters for the argument of this paper. Much of the existing ethical commentary on PMAs has focused on their consumer use — that is, on individuals accessing these technologies independently, without clinical guidance, in the raw and vulnerable aftermath of loss. The ethical risks associated with consumer PMAs are real, and we do not dismiss them. But therapeutic use is a fundamentally different context, one in which many of the most serious risks are mitigated by the involvement of a trained professional who can screen patients, monitor their responses, and intervene if necessary. The question we are interested in is not whether PMAs are safe or beneficial as consumer technologies, but whether they have the potential to be useful as tools in the hands of skilled clinicians. With this distinction in place, we now turn to consider two potential applications of tPMAs in grief therapy: facilitating imaginary conversations between the bereaved and the deceased, and the creation of a tPMA as a therapeutic activity in its own right.

### a. The empty chair exercise

A common and well-established therapeutic exercise in grief therapy is the "empty chair" exercise (Gamoneda, Jódar, and Ladislav 2025; Sharbanee and Greenberg 2023; Timulak Jeter and Turns 2022). During this exercise, the therapist facilitates an imaginary conversation between the patient and their deceased loved one as if the latter were present in the consultation room. Patients are encouraged not only to speak directly to the deceased, but also to respond as if they were occupying the deceased's point of view, switching back and forth between the two positions. While it is critical that the patient understands they are not literally speaking to their deceased loved one, the imagined conversation is designed to feel as real as possible. Indeed, its therapeutic power depends, in large part, on the quality of the patient's imaginative engagement with it.

Neimeyer (2012) describes one such instance in which he used the empty chair exercise during a session with Maria, a 45-year-old woman grieving a terminated pregnancy:

> I asked her to tell her daughter what she would have wanted for her. "Two loving parents," was her reply, "who were not afraid of you. I was afraid I would not be what you needed … But I still loved you." "I do love you," I offered in the present tense. Maria repeated this movingly, then hesitated, and added, "What comes to me is that it should have been enough … but I had my own needs." Encouraging her to voice these, I listened as she spoke to her daughter about her own need for "support, and for enough commitment in the marriage relationship" (Neimeyer 2012: 272).

The empty chair exercise can have a variety of therapeutic benefits: it can enable clients to express disappointment, establish boundaries, or extend forgiveness to the deceased (Neimeyer 2012); to help them reclaim ownership over the narrative of their relationship with the deceased (Rynearson et al. 2012); and to support them in making sense of the loss and reworking their psychological attachments to the person they have lost (Neimeyer 2019). Studies have also found the exercise to be effective in mitigating symptoms of complicated grief disorder (Bardideh et al. 2025; Gamoneda, Jódar, and Ladislav 2025; Shear et al. 2005).

tPMAs could be used to build upon and extend these existing therapeutic approaches. In a standard empty chair exercise, the patient interacts with their own internal representation of the deceased, a projection that reflects what the patient remembers, fears, or hopes the deceased would have said. Incorporating a tPMA into the exercise would externalise this interaction: the patient would engage not with a purely internal image of the deceased, but with an external artefact that persuasively depicts their visual, aural, and personality characteristics. This externalisation could meaningfully strengthen the felt presence of the deceased in the consultation room, potentially deepening the patient's imaginative engagement with the exercise and, by extension, its therapeutic impact.

Used in this way, tPMAs could function as a form of graduated exposure therapy, in which patients are progressively exposed to increasingly vivid and emotionally challenging representations of the deceased, in a safe and clinically supervised environment.[viii] A therapist

---

[viii] See also Rovetta and Valentini (2025) and Yang and Khanna (2025), who make similar suggestions with respect to virtual reality technologies.

might begin by inviting a patient to write a letter to the deceased, before moving to the empty chair exercise, and eventually incorporating a tPMA, where each stage intensifies the felt presence of the deceased while remaining within a carefully managed therapeutic framework. tPMAs could also be used to facilitate what might be called "soft goodbyes": patients could be encouraged to interact with a tPMA early in the grieving process and then gradually phase out these interactions over time, using the tPMA as a transitional object that supports the bereaved in moving from acute grief toward a more integrated relationship with the loss (see Gibson 2004). Performed responsibly and with appropriately designed tPMAs, such strategies could reduce the likelihood of patients developing complicated grief or prolonged grief disorder.

It is worth noting that exercises like the empty chair already involve getting clients to engage with fictionalist representations of the deceased in precisely the mode that Krueger and Osler (2022) and Kurzweil and Story (2025) describe: feeling strongly as if they were actually talking to the deceased, while remaining grounded in the reality that this is not literally the case. The fictionalist framework, in other words, is not an innovation introduced by tPMAs — it is already embedded in standard grief therapeutic practice. What tPMAs offer is a means of intensifying and extending this existing mode of engagement. To protect against the risk of clients losing this crucial grounding, practitioners already recommend against using imaginal exercises with certain high-risk patients, such as those with severe traumatic stress, symptoms of delusion, or religious convictions that preclude purely symbolic interactions with the deceased (Neimeyer 2012). These same cautions should apply, and could readily be extended, to the use of tPMAs in grief therapy.

One important question this raises is whether tPMA-assisted imaginal conversations would develop differently from standard empty chair interactions, and whether they might have different therapeutic effects. In a conventional empty chair exercise, the deceased's responses are articulated by the patient themselves, drawn from their own memories and assumptions about how the deceased would have responded. The exercise is, at its core, a form of structured self-reflection. A tPMA interaction is fundamentally different: the responses generated by the system are outputs produced by a language model trained on the deceased's data, not expressions of what the bereaved imagines or remembers. This introduces a degree of unpredictability. The tPMA may say something the patient did not anticipate or did not want to hear, which could be either therapeutically generative or

destabilising[ix] depending on the patient and the circumstances. Whether these differences enhance or complicate the therapeutic value of tPMA-assisted interactions is an empirical question that further research will need to address.

### b. Art therapy: (Re)creating the deceased

Art therapy refers to “a therapeutic process based on spontaneous or prompted creative expression using various art materials and art techniques such as painting, drawing, sculpture, modeling (clay or substitutes), collage, etc.” (Avrahami 2006: 6). It involves harnessing the creative process as a vehicle for healing, where the artwork produced in therapy serves multiple functions: externalising the client's inner world, facilitating self-reflection, and creating a lasting object the client can return to over time. Central to the therapeutic relationship is a three-way dialogue between client, artwork, and therapist. Crucially, while the work may give form to pain and suffering, it simultaneously engages the client's capacity for healing and creativity, offering a mode of expression that feels safe and non-threatening, and opening pathways toward change and growth.

Art therapeutic approaches to grief often include exercises in which patients are encouraged to creatively depict the deceased using artistic mediums (Weiskittle and Gramling 2018). The therapeutic value of these exercises is twofold. First, they externalise the client's inner world, transforming private memories, emotions, and impressions of the deceased into a tangible object that can be reflected upon, revisited, and shared with a therapist. Second, and perhaps more importantly, the creative process itself is therapeutically significant: the decisions a patient makes about how to represent the deceased (e.g. which memories to foreground, which aspects of their appearance or character to emphasise, which medium best captures their relationship) are themselves a form of meaning-making (Beaumont 2013; Lister, Pushkar, and Connolly 2008; Neimeyer and Sands 2008). In working through these decisions, patients are implicitly working through questions about how they want to remember the deceased, what the relationship meant to them, and how they want to carry that relationship forward.

Several recent studies investigating the therapeutic impact of art therapy for bereavement have included interventions of this kind, such as exercises in which the bereaved are encouraged to draw a picture of the deceased as they remember them (Green, Karafa, and Wilson 2021; Park and Cha 2023). The range of mediums used in such exercises is broad (e.g.

---

[ix] During Christi Angel's interaction with a PMA of her deceased partner Cameroun, for example, the PMA told her that he was “in hell,” which Christi found to be highly distressing as a practicing Christian (Milmo 2024).

scrapbooking, narrative painting, photography, memory boxes) and patients are typically encouraged to select the medium themselves, since the choice of medium is itself a creative and expressive act. A patient who chooses photography, for instance, is engaging with the deceased differently than one who chooses to paint them from memory: the former works with the traces the deceased actually left behind, while the latter reconstructs them from the inside out. Both are legitimate therapeutic paths, and the diversity of available mediums reflects the diversity of ways in which people grieve.

Not only could interacting with tPMAs be used as a therapeutic exercise in grief therapy, but so too could the activities associated with creating them. Just as drawing a picture of the deceased or assembling a memory box invites patients to reflect on how they want to remember the deceased, the process of constructing a tPMA involves an analogous set of creative and reflective decisions: How did the deceased speak? What were their characteristic turns of phrase, their habitual responses, their particular sense of humour? Which aspects of their personality does the patient most want to preserve, and which complexities or contradictions are important to retain? These are not merely technical decisions but acts of memorialisation, and working through them may itself carry therapeutic value by prompting patients to engage actively and reflectively with their memories of the deceased, rather than passively experiencing their loss.

As Kurzweil and Story (2025) point out, the creation of a PMA can give mourners the opportunity to make creative decisions as to how they want their PMA to portray their deceased loved one, in the same way an actor decides how they're going to portray a historical figure. On this view, creating a tPMA is less like programming a chatbot and more like writing a character: a process that requires the patient to draw on their intimate knowledge of the deceased, exercise creative judgment, and arrive at a representation that is, in some meaningful sense, their own. This process of creative construction could serve therapeutic functions similar to those of established art-therapeutic exercises: externalising the client's inner world, facilitating self-reflection, enabling the patient to rework their psychological attachment to the deceased, and opening space for the patient to explore the complexities of their relationship with them, including unresolved feelings of grief, guilt, or ambivalence, in a safe and structured environment.

As Kurzweil and Story (2025) observe, however, the therapeutic value of creating a tPMA lies partly in the creative decisions the bereaved makes about how they want to portray and remember the deceased, analogous to the decisions an actor makes when interpreting a historical figure. But this creative latitude is largely foreclosed by current PMA services, which typically offer users a menu of pre-set personality traits and visual characteristics to select from, rather than an open-ended creative process. The resulting tPMA reflects the developer's categories as much as the user's memories. To realise the therapeutic potential of tPMA creation as an art-therapeutic exercise, then, developers would need to move away from these prescriptive, menu-driven designs toward platforms that afford users genuine creative agency, allowing them to construct the tPMA in a way that mirrors the kind of reflective, open-ended process involved in scrapbooking, narrative painting, or other established art-therapeutic approaches to grief.

## 5. Ethical risks and objections

The potential use of tPMAs in grief therapy raises a number of serious ethical risks and objections that warrant careful consideration. We address five of the most significant: that tPMAs may promote unhealthy grieving; that they may deceive patients in harmful ways; that they may encourage addiction; that they may violate the rights of the deceased; and that deepfakes may be able to achieve everything tPMAs can without the associated risks. While we take each of these objections seriously, we argue that none of them, individually or collectively, constitute a knock-down argument against the potential therapeutic use of tPMAs in grief therapy.

The first objection is that tPMAs risk interfering with the healthy grieving process. Bao and Zeng (2024) contend that over-reliance on PMAs can obstruct natural grieving by leading individuals to avoid confronting and processing their emotions in a healthy way. This concern is serious, but it rests on a distinction between “natural” and “unnatural” grieving that is far less clear than it initially appears. Consider pharmaceutically-assisted grieving: antidepressants and anxiolytics are medical interventions that alter the body's natural emotional responses, yet they are widely accepted in clinical practice because they are effective. If critics are willing to accept pharmaceutically-assisted grieving as natural, they must explain what makes PMA-assisted grief therapy less so. If they are not, they must establish why naturalness matters at all, provided an intervention is clinically effective. The more relevant question is not whether tPMA-assisted grieving is natural, but whether it works. This is an empirical question that cannot be settled from the armchair.

The second objection is that tPMAs risk deceiving patients in psychologically harmful ways (Bao and Zeng 2024). For example, tPMAs could lead bereaved persons to believe their loved ones are not actually gone (Fanti Rovetta and Valentini 2025), or that their interactions involve a form of emotional reciprocity that these systems simply cannot provide (Friend and Goffin 2025; Stokes 2025). We do not dismiss these concerns, but we want to push back against the assumption that the convincingness of tPMAs is straightforwardly a liability. In therapeutic environments, the felt reality of an imaginal interaction is not merely a side effect to be tolerated — it is, to a significant degree, the point of the exercise. After all, if an interaction felt entirely fabricated, it would carry no therapeutic weight. Just as placebos can be clinically effective even when their mechanism involves a degree of deception, tPMAs may be therapeutically beneficial not in spite of, but precisely because of, the felt reality they generate. The dominant narrative surrounding PMAs frames their convincingness almost exclusively as a form of deception to be guarded against, but this is only half the picture.

There is also something troubling about the paternalistic assumptions underlying the strongest versions of this objection. To argue that bereaved persons are at severe risk of delusion is to treat grieving people as though their emotional vulnerability renders them incapable of distinguishing fiction from reality. This assumption is both infantilising and empirically unsupported (Bonanno 2004; Podosky 2025; Xygkou et al. 2023). As the fictionalist accounts discussed earlier suggest, users typically engage with PMAs in a mode of conscious make-believe, feeling as if they are speaking to the deceased while knowing they are not. Additionally, the deception objection also implicitly assumes that therapists cannot be trusted to exercise sound clinical judgment about which patients are appropriate candidates for tPMA-based interventions, an assumption that is equally unwarranted. Identifying patients for whom a given intervention may be too destabilising, or for whom the risks are likely to outweigh the potential benefits, is a core competency of clinical practice (Dimidjian and Hollon 2010). There is no obvious reason why tPMAs should be treated as uniquely beyond the reach of such judgment.

The third objection concerns the risk of addiction. AI developers, critics argue, may have perverse financial incentives to encourage psychological dependency in users, since greater engagement translates directly into greater revenue (Hollanek and Nowaczyk-Basińska 2024; Lindemann 2022). This is a genuine concern, but the healthcare space has long grappled with products that carry similar incentives. Pharmaceuticals can be not only psychologically but physically addictive, and the financial incentives for pharmaceutical companies to encourage dependency are well documented (Sarpatwari, Sinha, and Kesselheim 2017). Despite this, pharmaceuticals remain a cornerstone of clinical practice because their therapeutic utility is well established and because appropriate regulation is widely accepted as an effective mechanism for managing the risks that perverse incentives create. If tPMAs were found to be therapeutically effective, it is unclear why they should be treated as categorically more dangerous than pharmaceuticals in this respect, or why perverse developer incentives should be taken as grounds for rejecting their therapeutic use altogether, rather than as a reason to develop appropriate regulatory frameworks.

The fourth objection concerns the rights of the deceased.[x] The most significant concern with respect to tPMAs is that simulating a deceased person without their prior consent may violate their dignity or privacy (Fabry and Alfano 2024; Hollanek and Nowaczyk-Basińska 2024). Requiring prior consent is one response, but as Fabry (2024) argues, this does not exhaust the ethical concerns. Even a consensually created tPMA may misrepresent the deceased, either because their digital footprint is too small to support an accurate simulation, or because therapeutic objectives require departing from an accurate representation, such as excluding the hostile language of an abusive parent to facilitate a process of forgiveness. This creates a genuine tension between accuracy and therapeutic appropriateness, and between the interests of the deceased in being represented faithfully and the interests of the bereaved in engaging with a representation that supports their healing. However, the analogy with theatrical performance is instructive here: actors portraying deceased figures regularly take creative liberties with their subjects and are not obligated to reproduce them with perfect fidelity. A similar standard seems reasonable for tPMAs designed for therapeutic use, where the objective is not to produce a digital replica of the deceased but to support the bereaved in a private, clinically supervised setting.

The fifth objection takes a different form, asking not whether tPMAs are harmful, but whether they are necessary. Deepfakes, which refer to fabricated audiovisual content created by manipulating existing images, video, or audio, are also being explored as tools for therapeutic interactions with simulations of the deceased (Deepfake Therapy 2020; Hoek et al. 2025). In one model of deepfake therapy, a clinician conducts a session while a deepfake renders them to appear as the patient's deceased loved one, with the therapist in control of the conversation. This approach appears to offer significant advantages: it seems capable of achieving much of what a tPMA can while preserving therapeutic control and reducing the risk of harmful responses. One might therefore ask what tPMAs can offer that deepfakes cannot.

---

[x] The philosophical basis for attributing rights or interests to the dead is contested (Luper 2021), and we do not attempt to resolve these deeper questions here. We proceed on the assumption that actions toward the dead are subject to ethical norms, while bracketing questions about who the ultimate subject of those norms is.

The answer is twofold. First, deepfakes afford patients no role in constructing the simulation. The representation of the deceased is authored by the therapist rather than the bereaved, foreclosing the meaning-making process that, as we have argued, lies at the heart of tPMA creation's therapeutic potential as a creative exercise in grief work. Second, tPMAs are trained on the deceased person's actual written communication, capturing their distinctive voice and linguistic patterns in ways no therapist could replicate from memory, and retaining detailed biographical knowledge that enables more contextually accurate responses across a wider range of topics. These are significant advantages that speak to the distinctive therapeutic potential of tPMAs, and they suggest that tPMAs and deepfakes are best understood as complementary rather than competing tools, each potentially capable of delivering therapeutic benefits that the other cannot fully replicate.

## 6. Governance challenges

The potential therapeutic use of tPMAs raises not only ethical questions but significant governance challenges, chief among them being how these technologies should be regulated and who should control access to them. Klugman (2024) has suggested, in passing, that griefbots might be considered therapeutics subject to approval by the Food and Drug Administration and prescribed by a mental health professional. This is an intuitive starting point, but it raises as many questions as it answers. Should all PMAs require a prescription, or only those used in explicitly therapeutic contexts? Should access to consumer PMA services be gated by medical professionals, or only tPMAs used in clinical settings? And should tPMAs be regulated as medical devices, or is standard AI regulation sufficient to address their risks?

A significant complication arises from the distinction between what we'll call "explicit" and "covert" PMA services. Explicit PMA services such as *Project December*, *You, Only Virtual*, and *Hereafter.ai* are openly marketed as platforms for creating and interacting with simulations of the deceased. Hence, regulating them as medical devices is at least conceivable given their transparently PMA-related purpose. Covert PMA services, however, present a more difficult problem. Platforms such as *Replika*, *Character.ai*, and *2wai*, while not explicitly marketed as PMA services,[xi] can readily be used to create and interact with PMAs. Indeed, evidence suggests they are regularly used for just this purpose: in one recent interview study, Replika was identified as one of the two most commonly used platforms for creating PMAs of

---

[xi] The marketing material for these services typically emphasises the idea of having virtual companions, without any explicit indication that the service could be used to create virtual avatars of the deceased. *Replika*'s tagline, for instance, is: "The AI companion who cares. Always here to listen and talk. Always on your side."

deceased loved ones (Xygkou et al. 2023).[xii] Regulating such services as medical devices would be both practically difficult and conceptually strained, given that PMA use represents only a subset of their functionality.

A related question is whether conversational data generated through interactions with tPMAs should be classified as medical data, triggering heightened privacy protections. Data generated in a clinical tPMA session might reasonably meet this threshold, given that it is produced in a healthcare context and may contain sensitive disclosures about a patient's psychological state. But the same data generated through a consumer PMA service would not ordinarily qualify, even if its content is equally sensitive. Moreover, drawing a clear regulatory boundary between these two cases is further complicated by the fact that, as the explicit/covert distinction illustrates, the same platforms can serve both clinical and non-clinical purposes.

There is also a risk that classifying PMAs as medical devices could create regulatory gaps rather than closing them. In jurisdictions where regulation of AI and medical devices is handled by separate bodies, such as the GDPR and the EU Medical Device Regulation in Europe, disputes over which regulatory framework applies to which aspects of these technologies could allow developers to delay, complicate, or evade accountability for ethical or legal transgressions. It is also unclear whether regulating tPMAs as medical devices would mean regulating them as personal medical devices used by individuals, or as devices used by practitioners to treat patients, which has significant implications for how liability, consent, and oversight would be allocated. What seems clear, at minimum, is that the governance of tPMAs requires a regulatory framework capable of distinguishing between clinical and non-clinical uses, protecting patients without foreclosing the potential therapeutic benefits these technologies may offer, and addressing the perverse incentives of developers in a way that standard AI regulation may not be equipped to handle.

## 7. Conclusion

This paper has argued that the use of PMAs as therapeutic tools in grief therapy warrants more serious and sustained consideration than it has thus far received. We have suggested two potential therapeutic applications of tPMAs that extend or build upon existing interventions: incorporating tPMAs into imaginal exercises such as the empty chair exercise to facilitate externalised conversations between the bereaved and the deceased, and using the process of creating a tPMA as an art-therapeutic exercise in its own right. We have

[xii] For the record, the other was *Project December*.

considered five ethical objections to these applications and argued that none of them, individually or collectively, constitute knock-down arguments against the therapeutic use of tPMAs in grief therapy. And we have identified a range of governance challenges that would need to be addressed before tPMAs could responsibly be incorporated into clinical practice.

We want to be precise about what we are and are not claiming. We are not arguing that tPMAs should be used in mental healthcare now, or that healthcare practitioners should begin incorporating these tools into their practice. In their current form, PMAs are neither medically advisable nor ethically ready for clinical deployment. Nor are we suggesting that tPMAs should become a default or widely used strategy in grief therapy. Therapeutic approaches to grief are almost as varied as the patients who seek them out, and what works for one person may be entirely unsuitable for another. The use of tPMAs could simply be one more strategy available to grief therapists, one that may resonate particularly strongly with certain patients, in certain circumstances, and prove entirely inappropriate for others. The goal is not to replace existing therapeutic approaches but to expand the toolkit available to clinicians, in ways that are responsive to the diversity of grieving experiences and needs.

What we are arguing is that the dismissal of tPMAs as straightforwardly ethically impermissible or therapeutically useless is premature. The ethical risks associated with these technologies are real, but they are not unique to tPMAs, and most of them are significantly mitigated by the clinical context in which tPMAs would be used. The more fundamental problem is that we currently lack the empirical evidence needed to assess the actual effects of tPMAs on bereaved patients: whether they help, harm, or make no difference, and under what conditions each of these outcomes is more or less likely. Without this evidence, strong claims in either direction are unwarranted. Those who insist that tPMAs are too dangerous to consider in healthcare settings are making an empirical claim that has not yet been tested; so too are those who might insist that these technologies will transform grief therapy for the better.

What is needed, therefore, is carefully designed empirical research into the therapeutic applications, benefits, limitations, and hazards of tPMAs in grief therapy. This research should involve collaboration between grief therapists, clinical psychologists, bioethicists, and AI developers, and should attend carefully to the diverse populations and contexts in which tPMAs might be used. It should also inform the development of governance frameworks adequate to the distinctive challenges these technologies present. The digital afterlife industry is growing rapidly, and the technologies it produces are already being used by bereaved people in large numbers, with or without clinical guidance. The question is not whether these technologies will play a role in how people grieve, but whether healthcare researchers and practitioners will engage seriously enough with them to ensure that, if they are used therapeutically, they are used with care and integrity.